\documentclass[12pt]{iopart}
\usepackage{amssymb}
\usepackage{graphicx}% Include figure files
\usepackage{dcolumn}% Align table columns on decimal point
\usepackage{bm}% bold math
\usepackage{epsfig}
\newcommand{\beq}{\begin{equation}}
\newcommand{\eeq}{\end{equation}}
\newcommand{\beqa}{\begin{eqnarray}}
\newcommand{\eeqa}{\end{eqnarray}}
\newcommand{\ba}{\begin{array}}
\newcommand{\ea}{\end{array}}
\usepackage{color}

\begin{document}

\title[Finite temperature effects in two-mode bosonic Josephson
junctions]
{Finite temperature effects in two-mode bosonic Josephson
junctions}
\author{G. Mazzarella, L. Salasnich, and F. Toigo}
\address{Dipartimento di Fisica e Astronomia ``Galileo Galilei''
and Consorzio Nazionale Interuniversitario per la Scienze Fisiche della
Materia (CNISM), Universit\`a di Padova,
Via Marzolo 8, I-35131 Padova, Italy}

\date{\today}

\begin{abstract}
We analyze the effects of the temperature on a bosonic Josephson
junction realized with ultracold and dilute atoms in a double-well potential.
Starting from the eigenstates of the two-site Bose-Hubbard Hamiltonian,
we calculate the coherence visibility and the fluctuation
of the on-site occupation number and study them as functions
of the temperature. We show that, contrary to naive expectations,
when the boson-boson interaction is suitably
chosen thermal effects can increase the
coherence visibility and reduce the on-site number fluctuation.
\end{abstract}

\submitto{\JPB}
\maketitle

%%%%%%%%%%%%%%%%%%

\section{Introduction}

One-dimensional double-well potentials \cite{oliver} confining ultracold and
dilute bosons are the ideal arena to study the atomic counterpart of
the Josephson effect \cite{book-barone}. Bosonic Josephson junctions (BJJs) have been
widely explored at zero temperature. The lowest
energetic state of a BJJ is reasonably described by that of the two-site
Bose-Hubbard (BH) Hamiltonian \cite{milburn}. This Hamiltonian sustains
different ground-states depending on the coupling strength between the atoms. Then, varying the
interatomic interaction makes possible to engineer macroscopic coherent
states \cite{smerzi,stringa,anglin,mahmud,anna},
macroscopic Schr\"odinger-cat states \cite{cirac,dalvit,huang,carr,brand,cats}
and states characterized by high degrees of quantum correlations
\cite{cats,anna2}. The issue of which among the aforementioned
states emerges in a BJJ is addressed by diagonalizing the underlying Hamiltonian and
studying the (coherence) visibility of the interference fringes
in the momentum distribution of the bosonic cloud,
the quantum Fisher information, and the entropy entanglement
as functions of the boson-boson interaction \cite{cats}. It is worth to observe
that the coherence visibility gives a measure of the importance of the single
particle tunneling. On the other hand, the Fisher information is
related in a very simple way to the on-site number fluctuation \cite{dellanna} which is the natural quantity to
analyze the macroscopic quantum tunneling (MQT) of bosons across the
barrier. Incidentally, notice that the single particle tunneling occurs on
relatively short time scales, while the MQT takes place on times growing
exponentially with the number of bosons. The situation studied
within the two-site BH Hamiltonian framework is closely connected to
that of MQT between the two classical N\'eel configurations of a two-site spin $S$ easy axis
antiferromagnet Hamiltonian \cite{meier} which reasonably describes
antiferromagnetic molecular rings, as Fe$_{6}$ and Cr$_{8}$ \cite{lasc,parola}.
In realistic situations, however, quantum systems work at finite temperature.
For low dimensional geometries, effects due to thermal fluctuations are
important \cite{ft}. Studying these effects on atoms in a double-well is very important as discussed experimentally in \cite{gati}
and theoretically in \cite{fermidyn} for spin-polarized fermions and in \cite{weiss} for repulsive bosons.

In this work we extend the studies presented in \cite{cats} and
in \cite{weiss}. In particular, we investigate the coherence
and the on-site number fluctuation at finite temperature both in the attractive and in
the repulsive regime by using as theoretical tool the two-site BH Hamiltonian.
First we introduce the two-site BH model that we use
to describe the Bose-Einstein condensate in a
double-well. Then we diagonalize the BH Hamiltonian and study the
coherence visibility and the on-site number fluctuation as functions of the
interatomic coupling both at zero and at finite temperature. We point out that
in the attractive regime the coherence visibility of the system
exhibits an enhancement by increasing the temperature.
We explain this effect by analyzing the coherence visibility
of the thermally populated excited states. By using the Hellmann-Feynman theorem \cite{cohen} and the time independent perturbation theory in the deep attractive regime we obtain an analytical formula for the coherence visibility pertaining to the two lowest BH Hamiltonian eigenstates. This formula is used to describe the afore mentioned thermal enhancement. We perform a similar study for the on-site number fluctuation which exhibits a thermal softening in the attractive regime. Finally, we investigate the size effects on these quantities by varying the number of bosons in the system.

\section{The model Hamiltonian}

We consider $N$ identical interacting bosons of mass $m$ confined
by a trapping potential $V_{trap}({\bf r})$. This
potential can be realized by the superposition of an isotropic harmonic
confinement in the transverse radial plane and a double-well potential (DWP) $V_{DW}(x)$ in the axial direction $x$.
Then, $V_{trap}({\bf r})$ is given by
\beq
\label{trap}
V_{trap}({\bf r})=V_{DW}(x) +\frac{m\,\omega_{\bot}^2}{2}\,(y^2+z^2)
\;,\eeq
where $\omega_{\bot}$ is the trapping frequency in the radial plane.
We assume that the transverse energy $\hbar \omega_{\bot}$ is much larger than
the characteristic trapping energy along the $x$ axis due to the
potential $V_{DW}(x)$. Then, the dynamics of the system is quasi one-dimensional (1D).

In the following the system will be analyzed at finite temperature, $T>0$. We assume that both $k_BT$ ($k_B$ is the Boltzmann's
constant) and the characteristic boson-boson interaction energy are not much larger than the
gap between the two states of the lowest doublet of the double-well linear problem
\cite{fermidyn}, while they are much smaller than the gap between the first and the second doublet. In this case, only the two states in the lowest doublet will have non negligible occupancy.
If the two wells are symmetric, the system will be described by the
effective two-site Bose-Hubbard (BH) Hamiltonian \cite{milburn}
\beq
\hat{H} = -J\big(\hat{a}^{\dagger}_L\hat{a}_R
+\hat{a}^{\dagger}_R\hat{a}_L\big)
+\frac{U}{2}\big( \hat{n}_L (\hat{n}_L -1)
+ \hat{n}_R (\hat{n}_R -1)\big) \; .
\label{twomode}
\eeq
Here $\hat{a}_{k},\hat{a}^{\dagger}_{k}$ ($k=L,R$, where $L$ stays for left and $R$ for right)
are bosonic operators satisfying the usual commutation rules;
%algebra
%$[\hat{a}_{k},\hat{a}^{\dagger}_{l}]=\delta_{kl}$;
$\hat{n}_{k}=\hat{a}^{\dagger}_{k}\hat{a}_{k}$ is the number of
particles in the $k$th well; $U$
is the boson-boson interaction amplitude, and $J$ is the tunneling matrix
element between the two wells. The total number operator
$\hat{N}= \hat{n}_{L} +\hat{n}_{R}$ commutes with the Hamiltonian
(\ref{twomode}).

The spectrum of the Hamiltonian (\ref{twomode}) is determined by solving the
eigenvalues problem
\beq
{\hat H} |E_j\rangle = E_j |E_j \rangle
\eeq
for a fixed number $N$ of bosons. Since the Hamiltonian ${\hat H}$
preserves the total number of particles, it can be represented by a $(N+1)\times
(N+1)$ matrix in the Fock basis $|i,N-i\rangle$ ($i=0,...,N$). In this ket, the
left (right) index denotes the number of bosons in the left (right) well.
For each eigenvalue $E_j$ ($j=0,1,...,N$) the associated eigenstate
$|E_j\rangle$ will be of the form
\beq
|E_j\rangle=\sum_{i=0}^{N}\,c_{i}^{(j)} \, |i,N-i\rangle \; .
\label{eigenstate}
\eeq
We assume that the coefficients $c_{i}^{(j)}$ are real \cite{cats}.

The macroscopic parameters of the Hamiltonian (\ref{twomode})
are explicitly related to the atom-atom coupling
constant $g=4\pi\hbar^2a_s/m$ (with $a_s$ the s-wave scattering
length), to the atomic mass $m$ and the frequency $\omega_{\bot}$
of the harmonic trap (see, for example, \cite{ajj2}).
The on-site interaction amplitude $U$ is positive (negative) if $a_s$ is
positive (negative), so that it may be changed at will by Feshbach resonance.
Note that when $a_s<0$, to avoid the collapse, the system has to be
prepared in such a way that the $\displaystyle{|U| \lesssim \frac{0.4}{N}\hbar (\omega_{x}\omega_{\bot})^{1/2}} $, \cite{gammal}, with $\omega_{x}$ the trapping frequency of the single well.
The hopping amplitude $J$ is equal to $(\epsilon_{1}-\epsilon_{0})/2$,
where $\epsilon_{0}$ and $\epsilon_{1}$ are the ground-state and the first excited state energies of a
single boson in the double-well potential \cite{fermidyn}.

Note that at fixed number $N$ of bosons, the ground-state of the system depends on the
parameter $\zeta=U/J$ \cite{cats}. This parameter shall be used also throughout the present work to discuss the role of the boson-boson interaction in determining the system properties. In terms of $\zeta$ the previous condition about the collapse thus reads $\displaystyle{|\zeta| \lesssim \frac{0.4}{N J} \hbar(\omega_{x}\omega_{\bot})^{1/2}}$. Then, fixed $N$ and the radial frequency, the collapse can be avoided by suitably adjusting the parameters of the DWP.

\section{Analysis}

When the temperature is finite, the system is in a statistical mixture of
states. We shall work in the canonical ensemble. In the basis of the eigenstates of the Hamiltonian (\ref{twomode}) the mixed state density matrix
$\hat{\rho}$
is
\beq
\label{mixed}
\hat{\rho}=\sum_{j=0}^{N} \frac{e^{-\beta E_j}}{Z} |E_j\rangle\langle E_j|\;.
\eeq
Here $Z$ is the partition function given by $Z=\sum_{j=0}^{N}\ e^{-\beta E_j}$,
%\beq
%Z=\sum_{j=0}^{N}\ e^{-\beta E_j} \;,\eeq
where $\beta=1/(k_B T)$ with $k_B$ the constant of Boltzmann and $T$ the
absolute temperature. The thermal average of an operator ${\hat A}$ is then:
\beqa
\label{ta}
\langle {\hat A}\rangle
=\sum_{j=0}^{N} \frac{e^{-\beta E_j}}{Z}\langle E_j| {\hat A} |E_j\rangle \
\;,
\eeqa
which, in the zero temperature limit, gets back the expectation value
of $\hat{A}$ in the ground-state.

\subsection{Coherence visibility}

Recently, we have analyzed the coherence between the two wells of $V_{DW}(x)$ at zero temperature by calculating
the coherence visibility \cite{cats}.
At finite temperatures it is given by
\beq
\label{alphaT0}
\alpha_T=\frac{2}{N} \langle \hat{a}^{\dagger}_{L} \hat{a}_R \rangle \;,
\eeq
where the average $\langle...\rangle$ is the thermal average defined in
(\ref{ta}). Since we are assuming that the coefficients
$c_{i}^{(j)}$ are real (see Sec. II), $\langle \hat{a}^{\dagger}_{L}
\hat{a}_R \rangle\ =\langle \hat{a}^{\dagger}_{R} \hat{a}_L \rangle\ $. The operator $\hat{a}^{\dagger}_{L}
\hat{a}_R $ ($\hat{a}^{\dagger}_{R} \hat{a}_L$) destroys a single boson in the right (left) well and creates it in
the left (right) well. Then $\alpha_T$, see Eq. (\ref{alphaT0}), characterizes the single particle tunneling
through the barrier. The subsequent analysis will be then useful to
understand how the temperature affects the tunneling of single bosons.

\begin{figure}[ht]
\epsfig{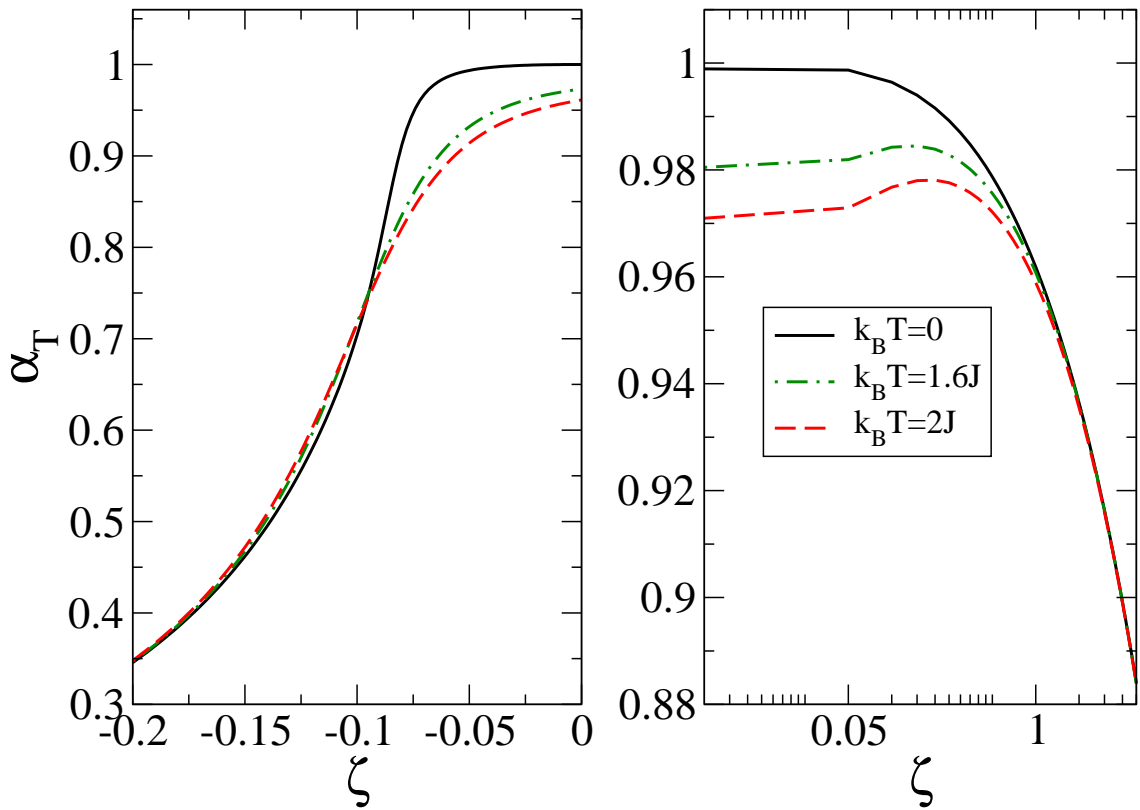}
\caption{(Color online). Coherence visibility $\alpha_T$ vs scaled interaction
strength $\zeta=U/J$ at $k_BT=0$ (solid line), $k_BT=1.6J$ (dot-dashed line), and $k_BT=2J$ (dashed line). Left panel: attractive
bosons, $\zeta <0$. Right panel (the horizontal axis is in logarithmic scale):
repulsive bosons, $\zeta >0$. Number of bosons $N=30$.}
\label{fig1}
\end{figure}

By using the expansion (\ref{eigenstate}) in Eq. (\ref{alphaT0}), $\alpha_T$
reads
\beq
\label{alphaT2}
\alpha_T =\frac{2}{N}\sum_{j=0}^{N} \frac{e^{-\beta E_j} }{Z}\sum_{i=0}^{N}
c^{(j)}_{i} c^{(j)}_{i+1}
\sqrt{(i+1)(N-i)} \; .
\eeq
Note that for $i=N$ the squared root factor is zero. Therefore the summation in the
right-hand side of Eq. (\ref{alphaT2}) is well defined even if
$c^{(j)}_{N+1}$ is not defined in Eq. (\ref{eigenstate}).

In Fig. \ref{fig1} we report $\alpha_T$ as a function of the parameter $\zeta=U/J$ both at zero and at finite temperature for $N=30$ bosons.  From the left panel, relative to the case of attractive bosons, we notice that, contrary to expectations, the coherence visibility at finite temperature is smaller than that at zero temperature only if the interaction strength is below a certain value. For interaction strengths larger than such a value the coherence visibility exhibits an enhancement over its zero temperature value. We have numerically verified that this behaviour occurs when $k_BT\gtrsim 0.5J$. As we shall show, this corresponds to values of $k_BT$ of the same order of the gap between the two lowest eigenstates of the Hamiltonian (\ref{twomode}).

On the contrary,  when the interactions are repulsive (right panel),  we find that the  coherence visibility  at finite temperature (dot-dashed and dashed lines) is smaller or equal to its zero temperature counterpart (solid line) for any value of the interaction strength, in agreement with the experimental findings of  Gati and co-workers \cite{gati}.

\begin{figure}[ht]
\epsfig{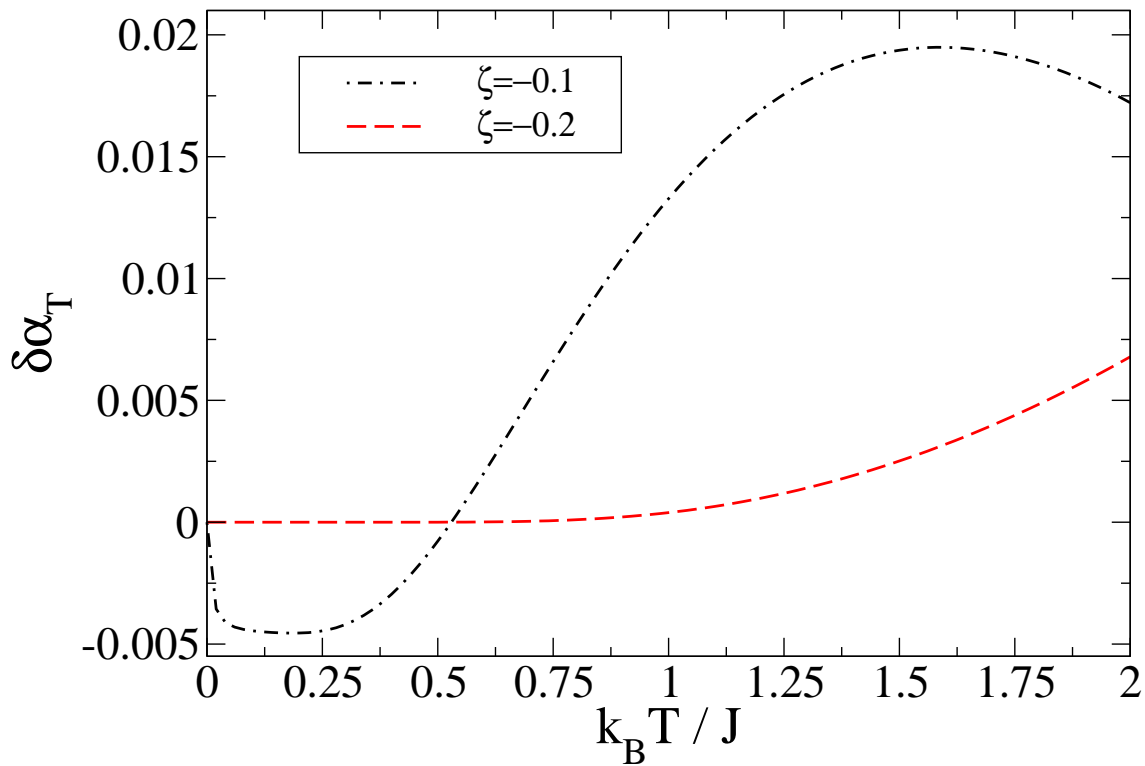}
\caption{(Color online). Relative variation $\delta
\alpha_T$ of the coherence visibility vs scaled temperature $k_BT/J$, calculated for scaled interaction strengths
$\zeta=-0.1$ (dot-dashed line) and for $\zeta=-0.2$ (dashed line).
Number of bosons $N=30$.}
\label{fig2}
\end{figure}

Due the unexpected behaviour of $\alpha_T$, we focus on the attractive regime and we investigate the relative change of the coherence visibility versus temperature for values of $\zeta$ for which the coherence thermal enhancement is expected to occur. In particular, we study the relative variation
\beq
\label{pva}
\delta \alpha_T=\frac{\alpha_T-\alpha_0}{\alpha_0}\;
\eeq
with $\alpha_0$ the coherence visibility at $T=0$ and
report the results in Fig. \ref{fig2}. When $\zeta=-0.1$
(dot-dashed line), $\delta \alpha_T$ is negative up to a given temperature, after
that it is positive. For $\zeta=-0.2$ (dashed line), the relative
change of the coherence visibility is always positive when the thermal energy is sufficiently high.
The positive values of $\delta \alpha_T$ for $\zeta=-0.1$ (dot-dashed line of the Fig. \ref{fig2}) are larger than those at  $\zeta=-0.2$ (dashed line of Fig. \ref{fig2}).
%This behavior can be related to the fact that at $T>0$ bosons may populate
%excited states of the Hamiltonian (\ref{twomode}) which have a larger coherence than the ground-state.

%\begin{figure}[ht]
%\epsfig{file=coeff2.eps,width=0.96\linewidth,clip=}
%\caption{(Color online). Coefficients $|c_{i}^{(j)}|^2$ for the four lowest
%eigenstates of Hamiltonian (\ref{twomode}),
%for $\zeta=-0.1$ (left panel) and for  $\zeta=-0.2$ (right panel). $\alpha_{j}$ is the coherence visibility of
%the $j$th eigenstate. Number of bosons $N=30$.}
%\label{fig3}
%\end{figure}

\begin{figure}[ht]
\epsfig{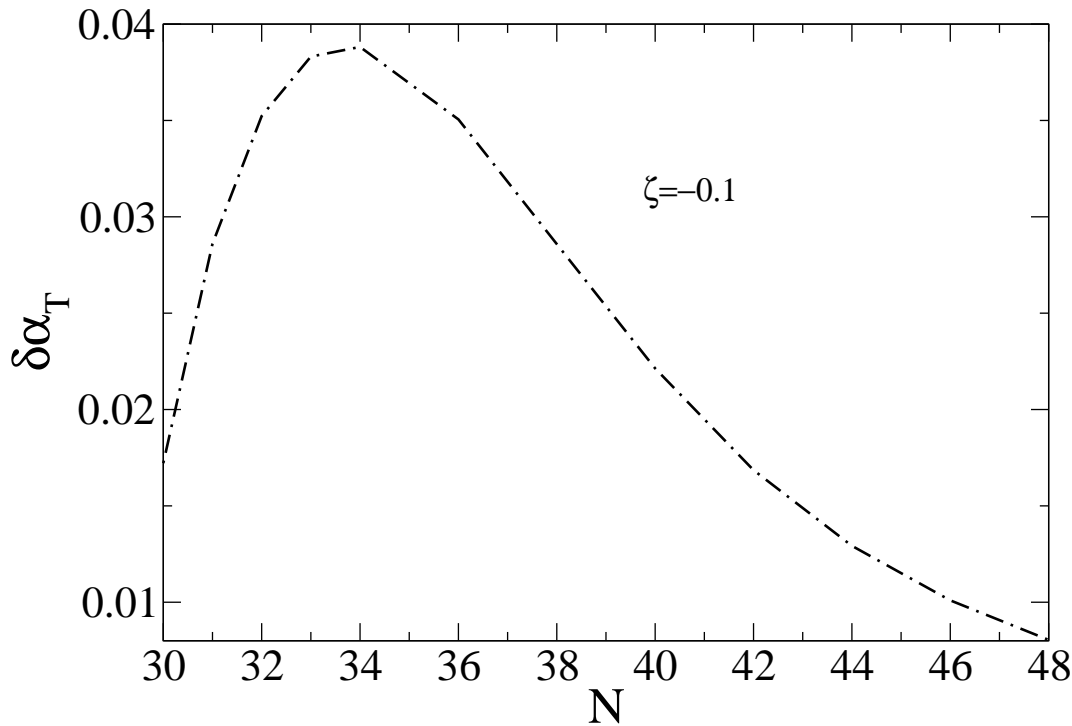}
\caption{Relative variation $\delta \alpha_T$ of the
coherence visibility vs number of bosons $N$,
calculated at $k_BT=2J$ for $\zeta=-0.1$.}
\label{fig4}
\end{figure}

%This interpretation is supported by the plots of Fig. \ref{fig3}. In this figure we show the
%quantities $|c_{i}^{(j)}|^2$ for the four lowest eigenstates of the Hamiltonian (\ref{twomode}) and the coherence visibility $\alpha_j$ for these states in correspondence to $\zeta=-0.1$ (left panel) and $\zeta=0.2$ (right panel).

The thermal enhancement of the coherence visibility can be related to the fact that at $T>0$ bosons may populate excited states of the Hamiltonian (\ref{twomode}) which have a larger coherence than the ground-state. To support this interpretation we use a simplified model in which we suppose that  $k_BT \simeq (E_1-E_0)$ so that only the two lowest eigenstates of the Hamiltonian (\ref{twomode}) are populated. At this stage, the thermal coherence visibility $\alpha_T$ may be written as
\beq
\label{visa}
\alpha_T=\frac{\alpha_0}{1+e^{-1}}+\frac{\alpha_1\,e^{-1}}{1+e^{-1}}
\;.\eeq
Here $\alpha_j=\langle E_j|\hat{a}^{\dagger}_{L} \hat{a}_R|E_j\rangle$ is the visibility of the $j$th excited state of the two-site BH Hamiltonian.
We, then, have to prove that the right-hand side of Eq. (\ref{visa}) is larger than $\alpha_0$, i.e.
\beq
\label{prove1}
\alpha_1 -\alpha_0 > 0
\;.\eeq
We evaluate $\alpha_1-\alpha_0$ by jointly exploiting the Hellmann-Feynman (HF) theorem \cite{cohen} and the time independent perturbation theory when $U/J \rightarrow -\infty$ which provides results in excellent agreement with those deriving from the exact diagonalization also for moderately small values of $|\zeta|$  as found in \cite{cats} for $N=30$. As observed in \cite{cats}, the HF theorem allows to write:
\beq
\label{hftc}
\alpha_j=-\frac{1}{N}\frac{\partial E_j}{\partial J}
\;,\eeq
so that:
\beq
\label{hftcd}
\alpha_1-\alpha_0=\frac{1}{N}\big(\frac{\partial E_0}{\partial J}-\frac{\partial E_1}{\partial J}\big)
\;.\eeq
In the limit $U/J \rightarrow -\infty$, the hopping operator in the Hamiltonian (\ref{twomode}) can be treated within the perturbation theory. We calculate $E_0$ and $E_1$ at the first non vanishing order, and get for $\alpha_1-\alpha_0$ the following expression:
\beq
\label{hftce}
\alpha_1-\alpha_0=-\frac{4\,(N+1)}{N\,(N-1)\,(N-3)\,\zeta}
\;,\eeq
which is positive when $N>3$ when the interaction is attractive.

We can explicitly estimate the temperature at which  thermal effects  affect the coherence visibility. This will happen when $k_BT \simeq (E_1-E_0)$.  From perturbation theory one has:
\beq
\label{gap}
E_1-E_0=\frac{2\,(N+1)}{(N-1)\,(N-3)}\frac{J^2}{U}+U\,(1-N)
\;.\eeq
Equating this value to $k_BT$, in the limit of $N \gg 1$ we obtain:
\beq
\label{condition}
J^2 \simeq \frac{(U\,N)^2}{2}+ \frac{U\,N\,k_B\,T}{2}
\;.\eeq
At temperature $T$, the interaction energy $-UN/2 \simeq k_BT$, so that we get
\beq
\label{condition2}
J \simeq k_B\,T
\;.\eeq

It is fair to remark that, when present, the increase of the coherence visibility over its zero temperature value is quite small and it depends on the number of bosons $N$. This is shown in Fig. \ref{fig4}, where we have plotted $\delta \alpha_T$ with $\alpha_T$ evaluated at  $k_BT=2J$ as a function of $N$ for $\zeta=-0.1$. The vanishing of $\delta \alpha_T$ with $\alpha_T$ when $N$ increases is confirmed also for higher temperatures and larger attractive strengths. From Fig. \ref{fig4}, moreover, we can see that the maximum gain in visibility ($\sim 0.04$) is reached when $N=34$. Such a moderate $N$ corresponds to a rather significant shot noise and experimentally it is necessary to gain sufficient statistics to demonstrate the contrast enhancement.

% Note that with this interaction strength for a number of particles larger than $48$, since the breaking of the left-right symmetry not the predictions deriving from the Hamiltonian are not
 %reliable (\ref{twomode}). The relative variation of $\alpha_T$ increases when the number of bosons is changed from $N=30$ to $N=34$. When $N>34$, the greater is the number of
% bosons the smaller is the afore mentioned relative increasing.
%When $\zeta=-0.2$ (dashed line), $\delta \alpha_T$ decreases by rising $N$.
% We expect therefore that the above described effect disappears as $N$ becomes larger.

Summarizing, we have found that for attractive bosons when both the temperature of the system and the boson-boson interaction become sufficiently large, the coherence visibility exhibits a thermal enhancement which, however, vanishes for large $N$.
Since the system under investigation is a closed system, its coherence properties are only determined by the interplay between the tunneling coupling energy (resulting from the overlap of the wave functions localized in the two wells) and the localization energy due to the the interaction between the particles. This has been shown in \cite{stringa} for the case of repulsive interactions and  can be seen also for attractive bosons by evaluating the expectation value $E$ of the BH Hamiltonian (\ref{twomode}) in the state \cite{cats}
\beq
\label{CS}
|QC\rangle=|CS\rangle_{L} \otimes |CS\rangle_{R}
\;,\eeq
where $|CS\rangle_{k}$ ($k=L,R$) is the coherent state \cite {glauber} describing the Bose-Einstein condensate in the $k$th well. Thus, the expectation value $E$ (up to a constant) reads
\cite{cats}
\beq
\label{e1}
E=\langle QC|{\hat H}|QC\rangle =
-N\,J\sqrt{1-z^2}\cos \phi +\frac{N^2\,U}{4}\,z^2\;,
\eeq
where $z=(N_L-N_R)/N$ is the fractional population imbalance between the two wells and $\phi=\theta_R-\theta_L$ is the phase difference between the two condensates. Under the hypothesis of sufficiently small population imbalance, the energy (\ref{e1}) becomes
\beq
\label{e2}
E=-E_J\cos \phi +E_c (\Delta n)^2 \;,
\eeq
where $\Delta n=(N_L-N_R)/2$, $E_J=NJ$ is the Josephson coupling energy, and $E_c=U$ is the localization energy due to the boson-boson interaction.
When both temperature and boson-boson attraction are sufficiently large, $E_J$ overcomes the localization energy $E_c$ by producing a visibility increasing with the temperature.

\subsection{On-site number fluctuation}

In  \cite{cats} we have also analyzed the quantum Fisher information (QFI) at zero
temperature. The QFI is used as a probe of the emergence of the NOON state
 - written as a symmetric combination of the Fock states $|N,0\rangle$ and
$|0,N\rangle$ - and is related to the on-site number fluctuation \cite{dellanna}.

In the following we shall investigate the thermal average of the squared of the
on-site number fluctuation $(\Delta \hat{n}_k)^{2}_{T}$  ($k=L,R$ as usual).
According to Eq. (\ref{ta}):
\beqa
\label{ft1}
(\Delta \hat{n}_k)^{2}_{T} &=&
\sum_{j=0}^{N} \frac{e^{-\beta E_j}}{Z}\,\langle
E_j|\hat{n}_{k}^{2}|E_j\rangle\nonumber\\
&-&\bigg(\sum_{j=0}^{N} \frac{e^{-\beta E_j}}{Z}\,\langle
E_j|\hat{n}_{k}|E_j\rangle\bigg)^2\
\;.\eeqa
The operator $\hat{n}_k=\hat{a}^{\dagger}_{k}\hat{a}_{k}$ counts the number
of bosons in the $k$th well. Given a state of the bosonic junction, the squared variance (\ref{ft1}) measures the deviation of the $k$th well population from its expected value, such a deviation being due to the particles tunneling between the two wells.
Then, $(\Delta \hat{n}_k)^{2}_{T}$ can be used to characterize the collective
transfer of bosons across the central barrier of $V_{DW}(x)$. In the following we investigate the thermal effects on this macroscopic quantum tunneling (MQT).

To fix the ideas, let us focus on the number fluctuation
in the left well, $k=L$. In terms of the coefficients of
the expansion (\ref{eigenstate}),
$(\Delta \hat{n}_L)^{2}_{T} \equiv f_T $ is given by
\beqa
\label{ft2}
f_T &=&
\sum_{j=0}^{N}\frac{e^{-\beta
E_j}}{Z}\,\sum_{i=0}^{N}(c_{i}^{(j)})^2\,i^2\nonumber\\
&-&\bigg(\sum_{j=0}^{N} \frac{e^{-\beta E_j}}{Z}\,
\sum_{i=0}^{N}(c_{i}^{(j)})^2\,i\bigg)^2
\;. \nonumber\\
\eeqa

\begin{figure}[ht]
\epsfig{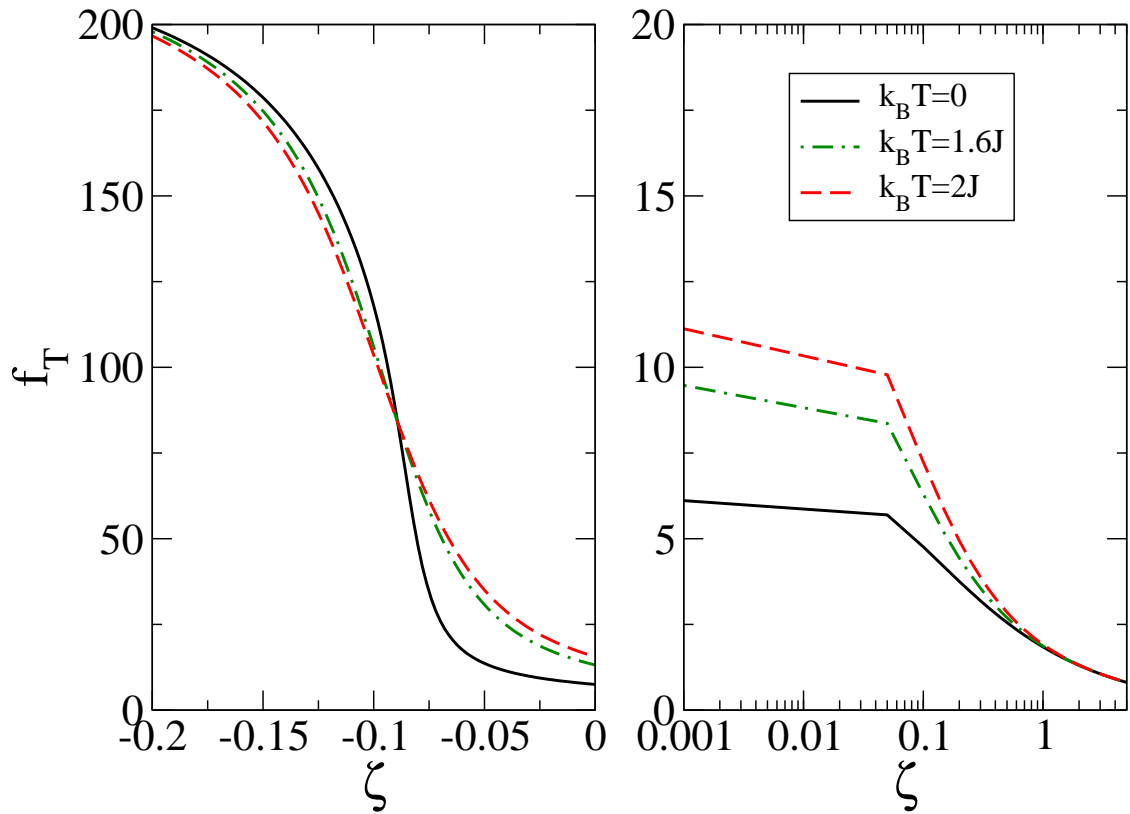}
\caption{(Color online). On-site number fluctuation $f_T$ for the left well population vs scaled interaction $\zeta=U/J$ at
$k_BT=0$ (solid line), $k_BT=1.6J$ (dot-dashed line), and $k_BT=2J$ (dashed line). Left panel: attractive
bosons, $\zeta <0$. Right panel (the horizontal axis is in logarithmic scale):
repulsive bosons, $\zeta >0$. Number of bosons $N=30$.}
\label{fig5}
\end{figure}

We have studied $f_T$ as a function of the scaled interaction parameter
$\zeta=U/J$ by varying the temperature, see Fig. \ref{fig5}. The solid line
of this figure represents the on-site number fluctuation at $T=0$, the dot-dashed and the dashed lines give the same quantity calculated,
respectively, at $k_B T=1.6J$ and $k_B T=2J$. From the plots of Fig. \ref{fig5}, we see that at finite temperature $f_T$ is enhanced over its zero temperature value for both attractive (for sufficiently weak boson-boson interactions) - left panel -  and repulsive bosons, right panel.
When the bosons are attractively interacting, it exists a value of $|\zeta|$, say $|\bar \zeta|$, above which the intra-well number fluctuations at $T>0$
are smaller than those at $T=0$. The strength of this interaction is very close to that for which the $f_T$ of the ground-state has a vanishing second derivative with respect to $\zeta$. It is interesting to observe that when $T=0$ and the interaction strength approaches $|\bar \zeta|$, the system begins to lose its coherence and the junction evolves towards the self-trapping regime \cite{cats}.

\begin{figure}[ht]
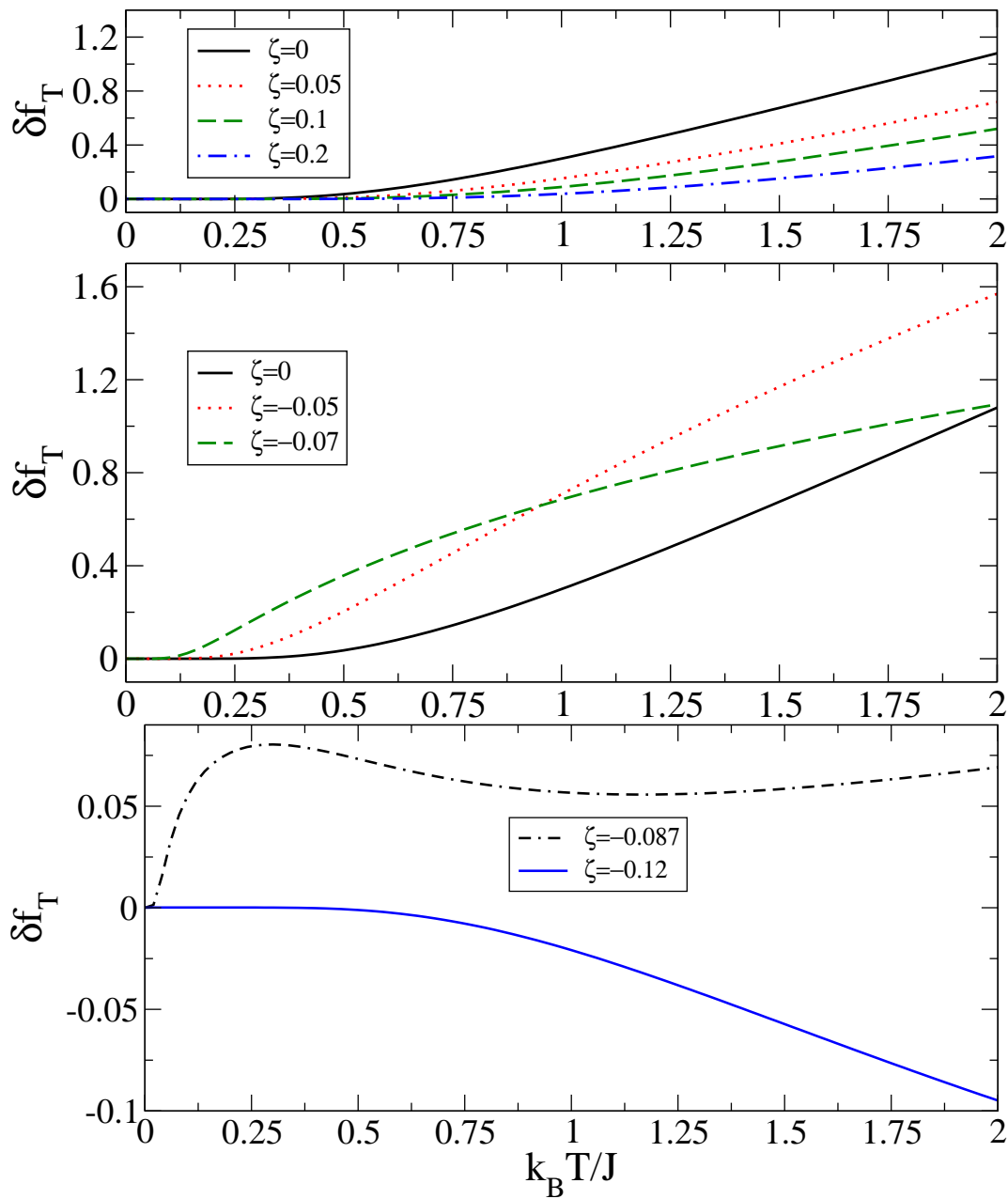

\epsfig{file=f5.eps,width=0.9\linewidth,clip=}
\epsfig{file=f5c.eps,width=0.9\linewidth,clip=}
\caption{(Color online). Relative variation $\delta f_T$ of
number fluctuation vs scaled temperature $k_B T/J$. Number of bosons $N=30$.}
\label{fig6}
\end{figure}

\begin{figure}[ht]
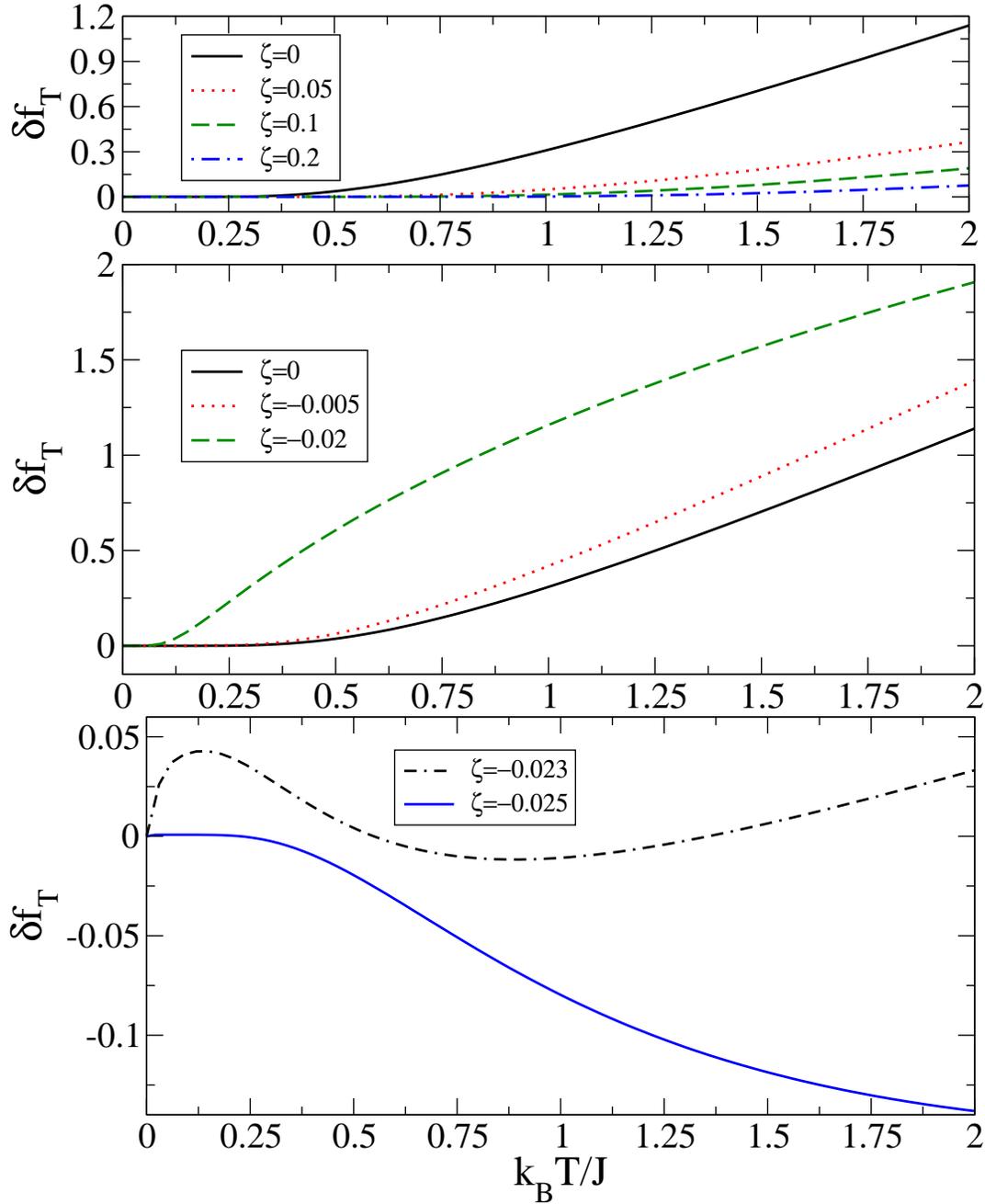

\epsfig{file=f6.eps,width=0.9\linewidth,clip=}
\epsfig{file=f6b.eps,width=0.9\linewidth,clip=}
\caption{(Color online). Relative variation $\delta f_T$ of number fluctuation
vs scaled temperature $k_B T/J$. Number of bosons $N=100$.}
\label{fig7}
\end{figure}

We now analyze the relative change of $f_T$ both in the attractive and in the repulsive regime. To this end, we
study $\delta f_T$ defined by:
\beq
\label{pvf}
\delta f_T=\frac{(\Delta \hat{n}_L)^{2}_{T}-(\Delta
\hat{n}_L)^{2}_{0}}{(\Delta \hat{n}_L)^{2}_{0}}\;
\eeq
as a function of the temperature in correspondence to different
interaction strengths. We have carried out this analysis
for $N=30$ and $N=100$, see Figs. \ref{fig6} and \ref{fig7}.

On the repulsive side (top panels of Figs. \ref{fig6} and \ref{fig7}), the relative change (\ref{pvf}) increases with the temperature.
The influence of the interatomic interaction is to reduce such a growth with the increasing of $\zeta$.
The middle and bottom panels of Figs. \ref{fig6} and \ref{fig7} are obtained for attractive bosons. From the middle panels (see the dashed lines therein) we observe that when the boson-boson attraction becomes sufficiently strong, a change in the concavity of
$\delta f_T$ takes place. By looking at the bottom panels we see that a further increasing of the interatomic attraction produces a maximum and a minimum ($\zeta \simeq -0.087$ for $N=30$ and $\zeta \simeq -0.023$ for $N=100$) in $\delta f_T$ that disappear for sufficiently high values of $\zeta$. At this point ($\zeta \simeq -0.12$ for $N=30$ and $\zeta \simeq -0.025$ for $N=100$) $\delta f_T$ is zero up to a given temperature, after that it decreases against $k_BT/J$, as shown by the solid lines of bottom panels of Figs. \ref{fig6} and \ref{fig7}. From these solid lines, it can be observed that the greater is $N$ the weaker is the attraction at which the MQT thermal softening sets in.

To explain the thermal softening of $f_T$ we follow the same path of reasoning used for the coherence visibility in the previous subsection . We suppose that  $k_BT \simeq (E_1-E_0)$ so that only the two lowest eigenstates of the Hamiltonian (\ref{twomode}) are populated. The thermal on-site number fluctuation  $f_T$ may be written as
\beq
\label{flz}
f_T=\frac{f_0}{1+e^{-1}}+\frac{f_1\,e^{-1}}{1+e^{-1}}
\;,\eeq
where $f_j=\langle E_j|(\Delta n_k)^2|E_j\rangle$. We, then, have to find the conditions under which the right-hand side of Eq. (\ref{flz}) is smaller than $f_0$, that is
\beq
\label{prove2}
f_1-f_0 < 0
\;.\eeq
To calculate $f_1-f_0$ we use, again, the time independent perturbation theory and the HF theorem to get \cite{cats}
\beq
\label{hftf}
f_j=\frac{\partial E_j}{\partial U}+\frac{N}{2}\,(1-\frac{N}{2})
\;.\eeq
and therefore:
\beq
\label{hft2}
f_{1}-f_{0}=\frac{\partial (E_1-E_0)}{\partial U}
\;.\eeq
In the deep attractive regime, we can treat the hopping operator within the perturbation theory and calculate  $E_0$ and $E_1$ to the first non vanishing order
%\beq
%\label{f0}
%f_0=\frac{N}{4}\bigg(N-\frac{4}{(N-1)\,\zeta^2}\bigg)
%\;,\eeq
to get $f_{1}-f_{0}$:
\beq
\label{hft3}
f_{1}-f_{0}=-\bigg(N+1+\frac{2\,(N+1)}{(N-1)\,(N-3)\,\zeta^2}\bigg)
\;,\eeq
which is negative for $N>3$.

%By using the Eqs. (\ref{f0}) and (\ref{hft3}), it is possible to show that given a $\zeta <0$ the condition
%(\ref{hft2}) is met when $N$ is suitably chosen.}
%Another
%For attractive bosons, in the $|U|N \rightarrow \infty$ limit, it is possible to find an analytical formula for the gap $E_1-E_0$
%in excellent agreement with the results of the BH Hamiltonian diagonalization \cite{cats}.
%It can be shown that the mentioned gap has a maximum for a given interaction
%strength, below which - see  Eq. (\ref{hft2}) - $f_1<f_0$. We may conclude that provided the interatomic interaction is
%sufficiently strong, a thermal depletion of $f_T$ is expected. Note that the absolute
%value of $U$ signing the maximum of $E_1-E_0$ decreases by increasing
%the number of bosons in the system.

\begin{figure}[ht]
\epsfig{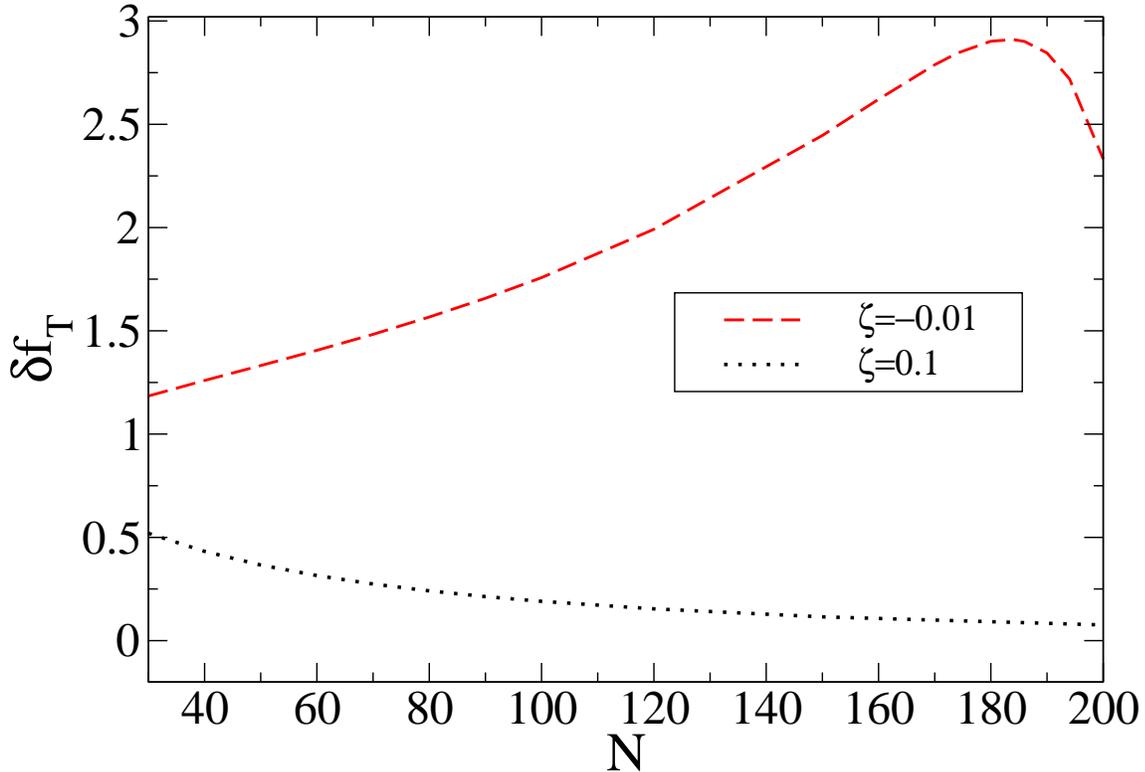}
\caption{(Color online). Relative variation $\delta f_T$, of the number fluctuation,
vs number of bosons $N$, calculated at $k_BT=2J$. Dashed line:
$\zeta=-0.01$. Solid line: $\zeta=0.1$.}
\label{fig8}
\end{figure}

\begin{figure}[ht]
\epsfig{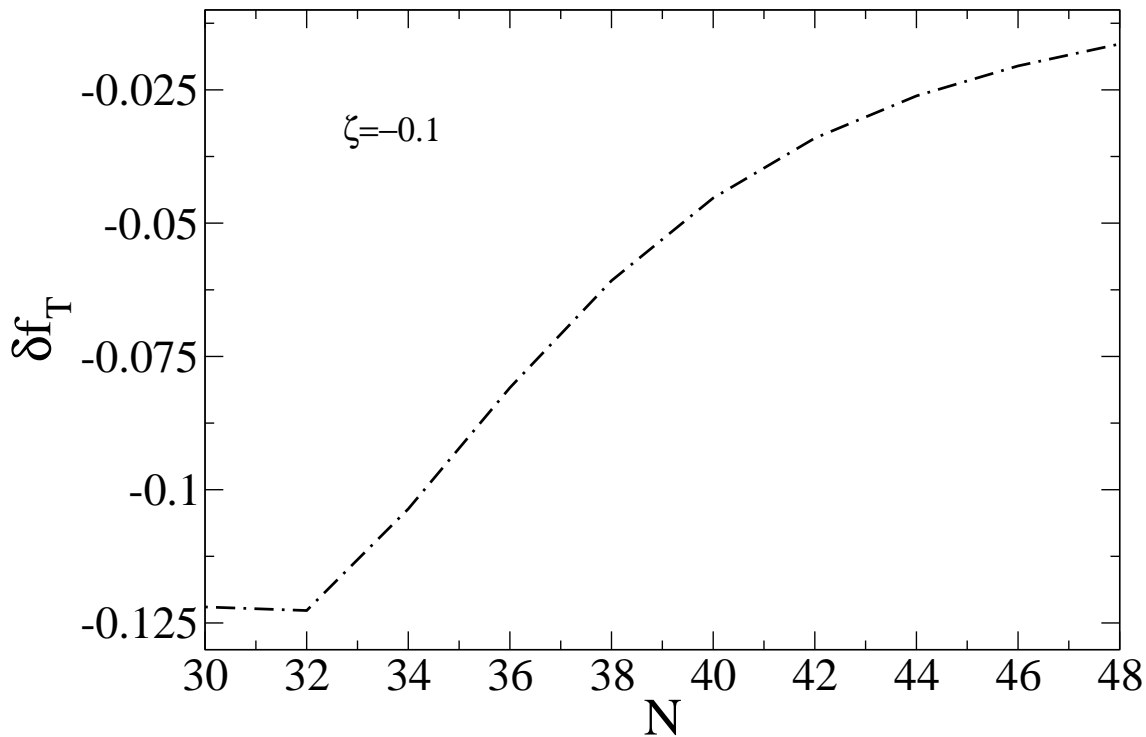}
\caption{Relative variation $\delta f_T$, of the number fluctuation,
vs number of bosons $N$, calculated at $k_BT=2J$ for $\zeta=-0.1$.}
\label{fig9}
\end{figure}

We have also analyzed the influence of the number of bosons both on the
thermal enhancement and on the thermal softening of the on-site number fluctuation.
Let us start with the case $\delta f_T>0$ shown in Fig. \ref{fig8}. It can be pointed out that
when the bosons interact attractively (dashed line) the relative
variation of $f_T$ increases with $N$ by reaching a maximum, after that it decreases at least up to the number of bosons that we have considered. On the other hand, for repulsive bosons (dotted line), $\delta f_T$ decreases by increasing $N$; in correspondence to the same number of
particles, the thermal enhancement of $f_T$ is more important for attractive
bosons.  The case $\delta f_T \le 0$ is reported in Fig. \ref{fig9}. It can be
observed that $\delta f_T$ increases by increasing the number of particles in the system.
%In Figs. \ref{fig8} and \ref{fig9}, the maximum number of bosons is that for which the left-right symmetry is broken and our model does not provide reliable predictions.

Let us, now, write down some conclusive remarks. The new predicted effects - coherence thermal enhancement and on-site fluctuation number thermal softening - would manifest at temperatures of about $J/k_B$ (see Fig. 2 and Fig. 6).
%In standard ultracold atom experiments in which optical lattices are employed to generate double-well potentials,
%hopping amplitudes such that $J/\hbar \simeq 10$ kHz are achievable \cite{oliverprv}. This results in temperatures of about $80$nK  which are within the experimental reach.
In the experiments performed to detect the scaling behavior of the coherence factor in a bosonic Josephson junction, the temperature varies in the range: $50$nK $<$ $T$ $<$ $80$nK, see, for example, \cite{gati}. It is fair to note that to observe the two mentioned phenomena in the above temperatures range, very large values of the hopping amplitude (corresponding to $V_{DW}$ with very low central barriers) - $6.6$kHz $<$ $J/\hbar$ $<$ $10.5$kHz - would be required.
%Then, to observe the discussed phenomena will become, eventually, possible only in the presence very large tunneling amplitude, that is with double-well potentials characterized by extremely low central barriers.

Possible limitations to the feasibility of the scenario analyzed in the present work are related to the assumption of the thermal equilibrium that we have made, see Eq.(\ref{mixed}). Only a sufficiently weak coupling between the confined bosons and an external thermal bath does not affect the physical properties of the system. For this reason, we expect that the coherence thermal enhancement and on-site fluctuation number thermal softening may be observed in mesoscopic systems only if they are very weakly coupled to the environment. When this condition is not met, the effect of the number fluctuation has to be considered, in particular for a moderately small number of bosons. This  problem is currently under study.

\section{Conclusions}

We have considered ultracold and dilute bosons in a one-dimensional double-well potential. By considering such a system at finite temperature, we have studied the thermal effects both on the single particle and on the macroscopic quantum tunneling. We have carried out this analysis by studying the coherence visibility and the on-site number fluctuation as functions of the temperature for different interaction strengths. We have pointed out that the thermal effects can increase the coherence visibility and reduce the on-site number fluctuation when the interatomic interaction is suitably tuned. We have explained the coherence thermal enhancement by analyzing the coherence visibility of the thermally populated excited states. By employing the Hellmann-Feynman theorem and the time independent perturbation theory, we have explicitly evaluated the coherence visibility pertaining to the two lowest eigenstates of the two-site Bose-Hubbard Hamiltonian.
A similar approach was been followed to calculate the on-site number fluctuations of the lowest states of the two-mode Hamiltonian and justify the observed thermal softening of the intra-well fluctuation. We have investigated size effects on the two mentioned quantities by varying the number of bosons in the system.\\

\section*{Acknowledgments}
The authors thank Massimo G. Palma, Oliver Morsch, and Alberto Parola for useful comments and discussions.\\

\end{document}